# Toxicity of lunar dust

Dag Linnarsson[1], James Carpenter[2], Bice Fubini[3], Per Gerde[4,5], Lars L. Karlsson[6], David J. Loftus[7], G. Kim Prisk[8], Urs Staufer[9], Erin M. Tranfield[10], and Wim van Westrenen[11,*]

[1]Section of Environmental Physiology, Department of Physiology and Pharmacology, Karolinska Institutet, Nanna Svartz v. 2, 171 77 Stockholm, Sweden, e-mail dag.linnarsson@ki.se

[2]European Space Agency ESTEC, HME-HFR, Keplerlaan 1, 2200 AG, Noordwijk, The Netherlands, e-mail james.carpenter@esa.int

[3]Department of Chemistry and Interdepartmental Center "G. Scansetti" for Studies on Asbestos and other Toxic Particulates, University of Torino, Via P. Giuria 7,10125 Torino, Italy, e-mail bice.fubini@unito.it

[4]Division of Physiology, The National Institute of Environmental Medicine, Karolinska Institutet, IMM, Box 287, 171 77 Stockholm, Sweden, e-mail per.gerde@ki.se

[5]Inhalation Sciences Sweden AB, Scheeles väg 1, 171 77 Stockholm, Sweden

[6]Section of Environmental Physiology, Department of Physiology and Pharmacology, Karolinska Institutet, Nanna Svartz v. 2, 171 77 Stockholm, Sweden, e-mail lars.karlsson@ki.se

[7]Space Biosciences Division, NASA Ames Research Center, Mail Stop 239-11, Moffett Field, CA 94301, USA, e-mail david.j.loftus@nasa.gov

[8]Departments of Medicine and Radiology, University of California, San Diego, 9500 Gilman Drive, La Jolla, CA 92093-0852, USA, e-mail kprisk@ucsd.edu

[9]Micro and Nano Engineering Laboratory, Delft University of Technology, Mekelweg 2, 2628 GR Delft, the Netherlands, e-mail u.staufer@tudelft.nl

[10]Cell Biology and Biophysics Unit, European Molecular Biology Laboratory, Meyerhofstrasse 1, 69117 Heidelberg, Germany, e-mail erin.tranfield@embl.de

[11]Faculty of Earth and Life Sciences, VU University Amsterdam, De Boelelaan 1085, 1081 HV Amsterdam, the Netherlands, e-mail w.van.westrenen@vu.nl

[*]Corresponding author, phone +31-20-3451390, fax +31-20-6462457

**Abstract**

The formation, composition and physical properties of lunar dust are incompletely characterised with regard to human health. While the physical and chemical determinants of dust toxicity for materials such as asbestos, quartz, volcanic ashes and urban particulate matter have been the focus of substantial research efforts, lunar dust properties, and therefore lunar dust toxicity may differ substantially. In this contribution, past and ongoing work on dust toxicity is reviewed, and major knowledge gaps that prevent an accurate assessment of lunar dust toxicity are identified. Finally, a range of studies using ground-based, low-gravity, and *in situ* measurements is recommended to address the identified knowledge gaps. Because none of the curated lunar samples exist in a pristine state that preserves the surface reactive chemical aspects thought to be present on the lunar surface, studies using this material carry with them considerable uncertainty in terms of fidelity. As a consequence, *in situ* data on lunar dust properties will be required to provide ground truth for ground-based studies quantifying the toxicity of dust exposure and the associated health risks during future manned lunar missions.

## 1. Introduction

The current renewed interest in human exploration of the Moon is driven not only by an urge to expand the human presence to other celestial bodies, but also by genuine scientific interest. Many aspects of the origin and evolution of the Earth and the other bodies in our solar system remain unclear. The Moon is thought to hold important information about the time when our own planet was formed, and humans remain capable of much more intelligent and adaptive exploration of the Moon than even the most sophisticated robotic and remote-controlled devices (e.g., Crawford et al., 2012). Identification and retrieval of representative or exotic mineral specimens, and drilling deep into the lunar subsurface are examples of tasks for which astronauts are superior to machines. The most compelling argument for human exploration is the unique ability of humans to identify and quickly assess the unexpected, enabling real-time adjustment of a pre-planned exploration strategy.

Although humans have landed on and returned from the Moon during the Apollo era, it is still a formidable challenge to secure the health and safety of astronauts during Moon missions. Challenges for future missions include long-term low- or microgravity, radiation exposure, and the maintenance of a number of life support systems during a much longer period than was the case during the Apollo flights (e.g., Cain, 2010, 2011).

One of the biggest challenges may be related to the presence of dust on the lunar surface. The ubiquity of fine dust particles on the surface of the Moon plays an important and often dual role in many aspects of human lunar exploration. On the one hand, identifying the mineralogical and chemical composition of the dust fraction of lunar soils can provide *in situ* geological context for both robotic and human landing sites. In addition, lunar dust may be an ideal starting material for a range of future *in* *situ* resource utilization activities on the Moon (e.g., Taylor et al., 2005), and dust is an important component of the lunar exosphere (Horanyi and Stern, 2011).

On the other hand, dust can adversely affect the performance of scientific and life-support instruments on the lunar surface. Fine dust was spread over all parts of the Apollo astronauts space suits, ending up in the habitat (Figure 1a), resulting in astronaut exposure times of several days. The Apollo astronauts reported undesirable effects affecting the skin, eyes and airways that could be related to exposure to the dust that had adhered to their space suits during their extravehicular activities, and was subsequently brought into their spacecraft (Figure 1b).

Dust exposure and inhalation could have a range of toxic effects on human lunar explorers, especially if longer exposure times become the norm during future manned exploration missions. There is therefore a need to assess the risks to health. The physical and chemical determinants of dust toxicity for terrestrial materials such as asbestos, quartz, volcanic ashes and urban particulate matter have been studied in great detail, and lunar dust simulant (synthesised from terrestrial volcanic material) has been found to exhibit toxic effects (Lam et al., 2002; Latch et al., 2008; Loftus et al., 2010). Unique features of actual lunar dust (described in more detail in section 3), resulting from its formation by (micro)meteoroid impacts and its extended radiation exposure in the absence of oxygen and humidity, could lead to toxic effects significantly exceeding those of simulants made from Earth materials. At present, the formation, composition and physical properties of lunar dust remain incompletely characterised with regard to human health.

In a micro-/hypo-gravity environment the risk of inhalation of dust is increased due to reduced gravity-induced sedimentation. Inhaled particles tend to deposit more peripherally and thus may be retained in the lungs for longer periods in reduced gravity as will be the case in a future lunar habitat (Darquenne and Prisk, 2008; Peterson et al., 2008). Inhalation of particles of varying size may affect the respiratory and cardiovascular systems in deleterious ways leading to airway inflammation and increased respiratory and cardiovascular morbidity (Frampton et al., 2006; Sundblad et al., 2002).

In this contribution, we review our knowledge of the physical chemistry determinants of dust toxicity, of the composition and size of lunar dust, and all aspects related to its toxicity. We identify a number of knowledge gaps that need to be filled to constrain the required extent of mitigation activities protecting astronauts from the potentially toxic effects of lunar dust during and after a stay on the Moon. We also recommend a range of future studies using ground-based, low-gravity, and *in situ* measurements on the lunar surface to better constrain lunar dust toxicity.

## 2. Physical chemistry determinants of dust toxicity

### *2.1. What makes a dust particle toxic?*

Up to the 1980s, fibrous character of asbestos (e.g., Kane, 1996; Mossman et al., 1990; Stanton et al., 1981), crystallinity of silica (e.g., Castranova et al., 1996; 1997; 2011), and degree of graphitization of carbon (e.g., IARC 1997; 2010) were considered the main physico-chemical determinants of the pathogenicity of these well-recognized particulate toxicants. Starting in the early 1990s, free radical release has been progressively accepted as a relevant additional factor in causing cell and tissue damage and DNA modifications (Fubini and Hubbard, 2003; Kamp and Weitzman, 1999; Sanchez et al., 2009; Shukla et al., 2003). The triggering or catalysis of these atomic-scale mechanisms by active sites located at the surface of the particles was subsequently elucidated (Fubini and Fenoglio, 2007; Fubini and Otero-Arean, 1999; Pezerat et al., 2007). Several additional aspects of surface chemistry and reactivity, including hydrophilicity / hydrophobicity and contamination by metals, are now also considered to play a role in particle toxicology.

With the advent of nanotechnology, there was an abrupt rise of interest in particle toxicology, because of the general fear that particles would exhibit an increased and/or new form of toxicity when nano-sized. Once again attention was directed to surface properties (e.g. extent and reactivity) as most of the adverse reactions caused by nanoparticles appear to take place at what has been designated as the "bionanointerface" (Fubini et al., 2011; Nel et al., 2009; Nel et al., 2006).

When toxicants act in particulate form, the mechanisms of toxicity are much more complex when compared to molecular toxicants, for the following main reasons:

- It is the *surface* of a particle which is in contact with fluids, cells and tissues.
- The same particle may act in *multiple stages* of the pathogenic process.
- The particle may stay *in vivo* for long periods of time, moving throughout the body.
- The particle may be *modified in vivo*.

The toxic potential of a particle is determined by several features rather than one specific property. It is generally accepted that three main factors contribute to the toxic potential of a given particulate: *form*, *surface reactivity* and *biopersistence* (Fubini and Otero-Arean, 1999). *Form* accounts for the fibrous character, nano-size, and complexity of particle morphology. *Surface reactivity* comprises all possible chemical reactions at the particle–living matter interface, which are determined or modulated by chemical composition, crystal structure, surface roughness (abraded versus smooth), origin (i.e. the way the material was prepared or reduced in particulate size) or contamination in the environment, e.g. by trace metals. Metal contamination may also occur *in vivo*, if the particle surface has potential to accommodate metal ions at surface charges or defects (Hardy and Aust, 1995a; Turci et al., 2011). *Biopersistence* is a term used to indicate the potential of a particle to remain substantially unaltered within a given biological compartment. It is related to dissolution in body fluids, but also to the efficiency of the methods of clearance, which both depend upon the shape of the particle, chemical composition and extent of the surface exposed. The longer a particle remains in a critical biological compartment causing stress to cells and tissues, the greater the risk of any adverse effect.

### *2.2. Fate of a particle in the body following inhalation*

A schematic view of the generally accepted mechanisms of toxicity of inhaled particles is presented in Figure 2. Note that the inflammatory reaction described here may occur via other penetration routes (e.g. skin, systemic circulation). Before interacting with cells, particles will adsorb molecules (mainly proteins) from the surrounding fluids, resulting in partial or full coating (Turci et al., 2010). This process depends upon both the particle surface properties and protein characteristics (Lynch and Dawson, 2008; Nel et al., 2009; Norde, 2008).

Naturally occurring antioxidants in the lung-lining fluids may react and be consumed at the particle surface (Figure 2, step 1) (Brown et al., 2000; Fenoglio et al., 2003; Fenoglio et al., 2000). Soon the immune system will engulf the foreign body by recruited alveolar macrophages (AMs) to clear the particle out of the lung (step 2). Depending upon the surface characteristics of the particle itself this clearance process may succeed (step 3) or fail (step 4). In the latter case, macrophages will become activated at the cellular and molecular level with activation of transcription factors and release of reactive oxygen or nitrogen species, chemotactic factors, lytic enzymes, cytokines, and growth factors (Fubini and Hubbard, 2003; Shukla et al., 2003; van Eeden et al., 2001). Eventually macrophages die (necrosis / apoptosis) and the particle is released at the lung surface again. Subsequent ingestion-re-ingestion cycles, accompanied by a continuous recruitment of alveolar macrophages, polymorphonuclear cells and lymphocytes, can cause sustained and chronic inflammation, which may evolve into several diseases, including cancer. Target cells such as bronchiolar and alveolar epithelial cells will then be affected by both alveolar macrophage products (step 5) and by direct interaction (step 6) with the extracellular particle itself (Fubini et al., 2010). Inhaled nanoparticles (NPs) partly escape alveolar macrophage surveillance following deposition in the alveolar regions of the lung, and may in some cases migrate to the pulmonary interstices or the systemic circulation (step 7) (Nel et al., 2009; Warheit et al., 2008), as discussed further in section 5.

### *2.3. Physical-chemical factors related to toxicity*

Free radical generation is the most strongly implicated feature in the mechanisms of toxicity. Free radicals, reactive oxygen species and transition metal ions act at stages 1, 3, 5 and 6 in Figure 2. Particle-derived free radicals associated to cell-derived reactive oxygen species and reactive nitrogen species cause oxidative stress in surrounding cells, which is exacerbated if the antioxidant defences are consumed (as shown schematically in Figure 3). When in contact with biological fluids, many dusts generate free radicals via various mechanisms, including reduction of oxygen, $OH^{\cdot}$ from hydrogen peroxide (Fenton mechanism) and homolytic rupture of carbon-hydrogen bonds (Fubini et al., 2010; Fubini and Hubbard, 2003).

Several surface states (i.e. surface sites which may exchange electrons) are associated with these reactions including “dangling bonds” (unsatisfied valences), poorly coordinated transition metal ions (typically iron or titanium), defects, and electron donating centres. Other surface properties relevant to toxicity are:

(1) *Surface charge and H-bonding potential.* These are the main determinants for phospholipid and protein adsorption, which in turn is implicated in cell adhesion and membrane activation or rupture (Nel et al., 2009; Nel et al., 2006).

(2) *Degree of hydrophilicity/hydrophobicity*, which determines the event following cell-particle interaction and clearance. Hydrophilicity originates in polar chemical functionalities (e.g. SiOH) or under-coordinated metal ions at the surface. For example, fully heat-hydrophobized cristobalite has been shown to lose its toxic potential towards several cell types (Fubini et al., 1999). The main roles of this property in the interaction of cell membranes with metal oxides have been described (Sahai, 2002). In the case of nanoparticles, hydrophobicity favours clumping and aggregation in aqueous media (carbon nanotubes are a typical example), whereas hydrophilic nanoparticles, aggregated in air, behave in a complex way in body fluids depending upon size and shape (Parsegian, 2006).

(3) *Dispersion of metal ions.* In addition to forming catalytic centres for free radical generation, well-dispersed metal ions may also be released in body fluids and tissues (Castranova et al., 1997; Hardy and Aust, 1995b).

(4) *Material association.* Association of different materials may result in new toxic entities, as shown for Indium Tin Oxide [ITO] (Lison et al., 2009) and for the combination of cobalt and tungsten carbide (Lison et al., 1995). In such cases, while the individual components are not, or only modestly toxic, their association generates a very toxic material. For example, cobalt + tungsten carbide particles are the cause of "Hard Metal Lung Disease" (Lison et al., 1995).

In the next section, we describe the composition and properties of lunar dust in light of the physico-chemical determinants discussed in this section, and assess their relevance to the potential toxicity of lunar dust.

## 3. Composition and size distribution of lunar dust

### *3.1. Characteristics of the lunar surface*

The surface of the Moon is covered by regolith. Regolith consists mainly of a wide range of crystalline and amorphous fragments produced by meteoroid impacts; glassy and partially crystallised spheres of volcanic origin, and agglutinates. Agglutinates is a term used for rock and mineral fragments welded together by glassy material formed upon impact of meteoroids (Lucey et al., 2006). Lunar soil generally refers to the size fraction of the regolith smaller than 1 cm in diameter (coinciding with the coarsest mesh size used to sieve Apollo samples upon return to Earth). The mean grain size of lunar soil varies between 40 and 800 µm, with an average at approximately 70 µm (Lucey et al., 2006). The term lunar dust is used for the fraction of lunar soil with a diameter smaller than 20 µm (McKay et al., 1991). For soil samples brought back to Earth during the Apollo missions, the proportion of dust in any soil is approximately 20 weight per cent [wt%] (Basu et al., 2001; Graf, 1993). Dust formation mechanisms, and more extensive overviews of dust properties, are provided in a recent review by Liu and Taylor (2011).

*3.2. Lunar dust and dust simulant properties*

Taylor and colleagues provided the most extensive studies to date of the mineralogical composition of two lunar dust fractions (sizes between 20 and 10 µm, and smaller than 10 µm, respectively) for nineteen different Apollo soils (Taylor et al., 2001a; Taylor et al., 2001b; Taylor et al., 2010). The samples were taken from both mare and highland areas of the lunar surface. Based on automated point-counting of > 150,000 areas in each sample, they conclude that > 70 % of dust consists of pyroxene, plagioclase, and impact glass-rich particles. The relative proportions of each phase vary as a function of bedrock type (mare versus highland), size fraction and soil maturity (a measure of the exposure age of the surface, with more mature soils having been exposed for longer periods of time). The more mature and smaller size fractions contain larger proportions of impact glass. Minor components of the dust include the minerals olivine and ilmenite as well as glass spherules formed in fire-fountaining volcanic eruptions on the lunar surface. A key observation, which may be important for lunar dust toxicity, is the presence of nanophase iron [np-Fe] in glassy rims of dust particles. This phase is formed during vapour deposition caused by flash heating of mineral or glass phases due to (micro)meteoroid impacts. The proportion of nanophase iron also appears to increase with decreasing dust grain size.

Based on the trends observed in the size fractions, Taylor and colleagues suggest that the respirable size fraction (typically considered to be < 5 µm) of lunar dust is likely dominated by impact glass and rich in metallic iron (Taylor et al., 2010). A recent study by Thompson and Christoffersen (2010) confirmed that approximately 80 % of submicron dust particles consists of glass. However, the proportion of np-Fe found was lower than predicted by Taylor and co-workers. This difference illustrates the first key issue with lunar dust studies in light of its possible toxicity: A comprehensive study of the properties and compositions of the very small size fraction of lunar dust (e.g., particles < 1 μm) is missing. Such a study will be required to bridge the current gap between size fractions relevant to toxicity studies and the currently state-of-the-art work by Taylor and co-workers, that groups all particles < 10 μm as the smallest fraction. Finally, we note that the Apollo samples studied to date are of near-equatorial origin and their mineralogy and physical properties may not be representative of other areas on the lunar surface including the South Polar region and the floor of the South Pole – Aitken basin, where future landing sites are proposed.

NASA developed several lunar soil analogues to minimize use of real lunar materials in mechanical engineering and *in situ* resource utilization studies. Lunar soil simulants JSC-1 (McKay et al., 1994) and JSC-1A (Hill et al., 2007) were produced from a Arizona volcanic tuff and ash deposit. JSC-1 is no longer available, and JSC-1A is commonly used as the lunar soil simulant. The mechanical properties of JSC-1A are close to those of real lunar soil. The simulant contains a large proportion (approximately 50 wt%) of volcanic glass, and its bulk chemical composition is similar to that of some of the Apollo 14 soils (Hill et al., 2007), which consist mainly of impact ejecta (which is not pure mare nor pure highland material). There are some key differences between JSC-1A and lunar soil. The glass in JSC-1A is not agglutanitic (was not formed due to flash heating during impacts) and is therefore not as friable. JSC-1A does not contain any ilmenite (a potentially toxic iron-titanium oxide that is present in all lunar soils); and it contains no nanophase iron. Instead, due to the presence of trivalent iron, nanoparticles of Ti-bearing magnetite are found in JSC-1A. These have magnetic properties mimicking those of nanophase iron (Hill et al., 2007; Liu and Taylor, 2011). Finally, JSC-1A contains approximately 0.7 wt% water incorporated mainly into clay minerals (McKay et al., 1994), whereas bulk lunar soil is virtually dry.

One lunar dust simulant of particular importance to lunar dust toxicity studies is JSC-1Avf, a < 20 μm sieved "very fine" fraction of JSC-1A. As far as we are aware, its mineralogical and chemical composition has not been studied in detail, but is assumed to be similar to that of JSC-1A (Hill et al., 2007).

Particle size is a key factor in toxicity studies because it defines transport pathways into the lung. Size reduction may also lead to an increase in surface reactivity (Fubini et al., 2010). Determining the size distributions of lunar dust particles was not a primary science objective at the time of Apollo and this important aspect for dust retention in the lungs should be reassessed. A summary of cumulative particle size distribution (PSD) measurements of lunar soils is given in Figure 4a, showing variations in the proportion of dust in Apollo soils between less than 10 wt% and greater than 30 wt%. Figure 4a illustrates a second knowledge gap regarding the toxicity of lunar dust: Classic PSD measurements did not consider particles smaller than 1 μm, and therefore 'miss' a portion of the respirable dust fraction. Careful analysis of the size distribution of the dust fraction of Apollo soils including the very fine fraction (Figure 4b) show log-normal number distributions with peaks in the 0.1 – 0.2 μm range (Park et al., 2008). This contrasts with analyses of JSC-1Avf, which peaks at approximately 0.7 μm. Finally, we note that although surface shape and surface area are key aspects in dust toxicity research, the surface morphology of lunar dust grains is at present poorly characterised (e.g. (Liu et al., 2008)). Electron microscope images of lunar dust show that many grains are highly vesicular suggesting a very high surface area, but quantitative data are currently absent.

*3.3. The potential toxicity of lunar dust*

Lunar dust appears to have many of the characteristics of particulate matter that cause adverse health effects (section 2). Hurowitz and colleagues observed that freshly fractured lunar regolith is able to produce large amounts of highly reactive oxygen species (Hurowitz et al., 2007) which are known to induce inflammation and adverse effects on cellular metabolism. Some peculiarities of lunar dust might trigger new toxicity paths and exacerbate the classical inhaled particle effects of terrestrial dusts. The surfaces exposed will all be generated by mechanical rupture of rocks and minerals, which causes a much higher surface reactivity (Damm and Peukert, 2009; Fenoglio et al., 2001) hence toxicity (Vallyathan et al., 1988) compared to unfractured material. Lunar dust particles will be rich in dangling bonds and unsatisfied valencies because of mechanical impacts and abrasion of bedrock and regolith (Figure 5). Furthermore on the Moon dangling bonds and charges will be more abundant and persist for much longer periods of time in the absence of oxygen or liquid water, which both assist defect healing and surface reconstruction (Wallace et al., 2010).

Vitreous materials, which are abundant on the lunar surface, have seldom been studied in terms of their toxicity. Their reactivity - hence toxicity - might be different from either their crystalline counterparts or other amorphous forms not obtained in particulate form by mechanical stress, but by e.g. sedimentation. This could have a large effect on toxicity, analogous to the observed toxicity contrast between vitreous and precipitated amorphous silicas (Ghiazza et al., 2010). Continuous exposure to radiation and solar winds will enrich the particles in reactive sites and electron donating centers. Photoactivation, similar to what takes place with titanium dioxide (Fenoglio et al., 2009; Fujishima et al., 2008), might also occur inducing the release of reactive oxygen species responsible for oxidative damage. This extremely large concentration of surface charges, unsatisfied valencies and reactive sites is expected to readily react when particles are immersed in any body fluid (Loftus et al., 2010; Wallace et al., 2010). All these features of lunar dust are consistent with the reports from Apollo astronauts on the extreme stickiness on all surfaces of the dust (e.g., NASA Manned Spacecraft Center, 1973).

As to the degree of hydrophilicity / hydrophobicity, it seems that fully hydrophobic (Fubini et al., 1999) and fully hydrophilic (Gazzano et al., submitted) silicas appear much less cytotoxic than specimens with mixtures of hydrophilic and hydrophobic sites. Lunar dust cannot be expected to show the same level of hydrophilicity as terrestrial dust due to the differences in hydrogen/water abundance between the Earth and the Moon. However, some surface active sites may readily react upon contact with water in the body. Under these circumstances, hydrophilic sites will appear at the particle surface but full hydroxylation will not been attained unless extremely long periods of time are available.

Nanophase iron particles are also an expected source of toxicity because of their size and the complex redox chemistry taking place at their surface when exposed to air. Such nanoparticles have been reported to be embedded into a vitreous matrix in lunar dust grain rims (Wallace et al., 2010). It is therefore unknown if direct interactions between the nanophase iron and the body will occur. Taking into account the long clearance times expected (which may allow partial dissolution of the amorphous silicate) and the persistent particle disruption taking place at the lunar surface, it seems likely that the iron surface might come in direct contact with cells and tissues. The toxicity of ilmenite - an iron titanium oxide - has never been considered and because of the presence of iron as well as titanium it might have adverse effects if inhaled.

These iron-containing materials could react with hydrogen peroxide to undergo a surface-induced Fenton-like reaction, ultimately producing reactive free radicals. Some similarities may be found between terrestrial volcanic ashes and lunar dust. In terrestrial ashes, iron is always only partially oxidised to trivalent iron. This appears to be one of the causes of sustained free radical generation, hence potential toxicity (Horwell et al., 2007; Horwell et al., 2010). Similarly, even after exposure to air lunar dust will keep a large fraction of its iron in a low oxidation state.

## 4. Transport of dust in the airways

### *4.1. Basic mechanisms of particle deposition*

The transport of inhaled particles in the airways is generally considered to be governed by three principal transport mechanisms: *inertial impaction* which primarily affects particles in a size range > ~5 µm; *sedimentation* which dominates the size range ~0.5 to 8 µm; and *diffusion* for particles smaller than ~0.5 µm (West, 2001). Of these, sedimentation is a gravity-driven process and so is altered in the lunar environment where acceleration due to gravity is only ~1/6 that on the Earth's surface.

As a consequence of these different transport mechanisms, the location in the bronchial tree where particles deposit also varies, and as a result, so does the time it takes to subsequently remove the deposited particles from the lung. Both the site of deposition, and the time for removal have the potential to affect the magnitude of the toxic effect of a given particle load delivered to the lungs, and so the changes in deposition in reduced gravity may alter the toxicological potential of an airborne dust. Thus large particles deposit primarily in the naso-pharangeal region (the nose and back of the throat) and do not reach the lungs. The medium-sized particles are generally considered to deposit in the small airways. Particles smaller than 100 nm in diameter deposit primarily in the alveolar region of the lung, with the exception of the very smallest particles (smaller than ~5-10 nm) which are primarily deposited in the nose (Geiser and Kreyling, 2010).

The importance of particle size in determining the health effects of dust exposure has long been appreciated. The U.S. Environmental Protection Agency (EPA) has more stringent regulations for particles smaller than 2.5 μm (termed $PM_{2.5}$, with a 24-hour exposure limit of 35 μg/m$^3$) compared to particles smaller than 10 μm (termed $PM_{10}$, a 24 hour exposure limit of 150 μg/m$^3$) (http://www.epa.gov/ttn/naaqs/standards/pm/s_pm_index.html). Early regulations controlled only total suspended particles (TSP), however this was replaced with a standard for $PM_{10}$ in 1987, and subsequently augmented with a standard for $PM_{2.5}$ in 1997, which was then revised downwards in 2006. Such changes reflect recognition of the importance of the smaller size fractions in impacting human health.

### *4.2. Deposition in reduced gravity*

Total deposition (the overall fraction of the inhaled particle load that deposits in the lung) and how that is altered by changes in the gravity level was studied in the 1990s in parabolic flight. While parabolic flight only provides short periods of reduced gravity (~25 seconds of microgravity), this is sufficient time for such studies. Parabolic flight also brings with it the advantage that the surrounding periods of hypergravity can also be utilized to extend the gravity range of the studies. These studies showed an expected reduction in deposition in reduced gravity; however, when the results were compared to predictions of deposition from existing models of aerosol transport in the lung, the observed deposition of particles of ~1 μm in size was considerably higher than expected in microgravity (Figure 6) (Darquenne et al., 1997). The implication drawn from these studies was that alveolar deposition of the smaller sized particles (0.5 and 1.0 μm) was higher than expected because of the combined effects of a higher alveolar concentration of particles in microgravity due to the absence of removal in the small airways by sedimentation, and the non-reversibility of the flow patterns in the complicated airway geometry.

The issue of non-reversible flow patterns in the lung leading to enhanced mixing between the inspired and resident gas is important in the context of aerosol deposition. While inspired gases rapidly and effectively mix with resident air through the process of molecular diffusion in the small acinar spaces of the lung, particles (even those small enough to have a significant diffusive transport component) mix much more slowly. Thus convective flow mixing is a potentially important mechanism by which particles transfer from the inspired air to resident air (and thus deposit). The higher than expected deposition of small particles seen in microgravity (Darquenne et al., 1997) implies that this transfer is more important than previously thought. It had long been assumed that because of the very low Reynolds number in the lung periphery, acinar flow was kinetically reversible. However, that concept is at odds with experimental data that show appreciable aerosol mixing at the alveolar level (Heyder et al., 1988) which occurs even in the absence of gravitational sedimentation (Darquenne et al., 1998, 1999).

A likely mechanism is the complex folding of streamlines that occurs in the bifurcating geometry of the airways. This chaotic mixing, termed “stretch and fold” because of its similarity to the repeated stretching and folding of a sheet of pastry, occurs due to the rapid stretching of flow streamlines that takes place as the cumulative airway cross-sectional area increases towards the lung periphery, and due to the slight asynchrony in the expansion and contraction of adjacent lung units (Butler and Tsuda, 1998; Tsuda et al., 2002). The effect is to bring streamlines into close apposition with each other such that diffusive transport of particles between inspired and resident air, that would otherwise be ineffective, now becomes an effective transport mechanism. Studies of aerosol mixing in microgravity incorporating small flow reversals have been performed in humans (Darquenne and Prisk, 2004), and observations of complex mixing patterns in rat lungs even after only one breath have been made (Tsuda et al., 2002). Together, these studies suggest that the higher than anticipated levels of both deposition and dispersion (a measure of transfer of inhaled particles to the resident air) seen in microgravity (Darquenne et al., 1998, 1999) result from the mechanism of stretch and fold.

### *4.3. Site of deposition*

The location of particle deposition in the airways has important implications for their potential toxic effects. The lungs are challenged by dusts on a continuous basis and as such particle deposition is an ever-present effect. The lung deals with this through clearance mechanisms. The medium and large-sized airway epithelium is covered in a layer of cilia, which serves to move a mucus layer in the direction of the glottis where it is swallowed. Particles that deposit on the mucus layer are transported by it and thus eliminated. This mucociliary clearance system is quite effective with clearance times in the order of hours and typically less than 24 hours for particles depositing in the central airways (Scheuch et al., 1996; Smaldone et al., 1988; Stahlhofen and et al., 1989). Hence, the mucociliary clearance system serves to reduce the time a potentially toxic particle can effect the underlying tissues (Fanizza et al., 2007).

In contrast, particles that reach the alveolar region of the lungs are removed by the alveolar macrophages, which engulf the particles (phagocytosis) and eventually transport them to the mucociliary clearance system for removal (Figure 2). However this process is much slower. A study in humans using magnetically-labelled particles showed that while approximately half the particles were removed from the lung with a mean residence time of $3.0 \pm 1.6$ hours, the remaining half of the particle burden had a mean residence time of $109 \pm 78$ *days* (Moller et al., 2004), greatly increasing the potential for toxic effects. In addition, Oberdörster and co-workers showed that nano-sized particles can escape macrophage surveillance, leading to even longer residence times than observed for micron-sized particles (Oberdörster et al., 2005b).

To date there are no published reports of human studies that provide specific information on how the site of deposition of inhaled particles is affected by altered gravity. However indirect measurements strongly point to a more peripheral site of deposition in low gravity, consistent with the concept of the reduction in sedimentation permitting higher particle concentrations reaching the alveolar region, where other mechanisms of deposition (diffusion, complex mixing) dominate. The only published study of aerosol transport in lunar gravity (Darquenne and Prisk, 2008) showed that while lunar gravity reduced overall deposition, for a given deposition fraction deposition occurred much more peripherally than in 1 G (Figure 7). The same study also showed that the unexpectedly high deposition of small particles (0.5 and 1.0 μm) seen in microgravity was still present in a lunar-gravity environment. The latter point is important in that it suggests that the results from previous studies performed in microgravity can be extrapolated to low-gravity, and that the mechanisms responsible for the higher than anticipated deposition are still operative in the presence of gravity. Recent studies have begun to measure the rate of clearance of particles deposited in humans in microgravity and compare that to the rate of clearance when the deposition occurs in 1 G (Prisk et al., unpublished observations). At present no results are available, but the expectation is that clearance will be slower as a result of the more peripheral deposition in microgravity.

There are now recent data on the changes in the site of deposition in rat lungs when gravity is altered. Rats were exposed to 0.9 µm particles containing ferric oxide in either 1 G or in microgravity during parabolic flight. The lungs were subsequently excised and imaged using MRI. When the particle exposure occurred in microgravity, the ratio of deposition in the central lung regions compared to the peripheral lung regions was greatly reduced ($0.58 \pm 0.20$ in microgravity; $1.06 \pm 0.08$ in 1 G; $P < 0.05$) showing that particles deposited more peripherally in microgravity (Darquenne et al, unpublished observations).

## 5. Non-pulmonary effects

### *5.1. Cardiovascular effects*

The cardiovascular system is directly linked to the pulmonary system at the blood-air interface in the lungs. Not only are gases able to cross this interface, but dust particles smaller than 100 nm are thought to cross from the alveolar surface into the pulmonary capillaries (Nemmar et al., 2002). Nanoparticles have been found in the lymph nodes (Brain et al., 1994; Harmsen et al., 1985), spleen (Semmler et al., 2004), heart (Semmler et al., 2004), liver (Oberdörster et al., 2000; Peters et al., 2006) and even the bladder (Nemmar et al., 2002) and brain (Oberdörster et al., 2004). Experimental evidence indicates that in exposed rats some nanoparticles reached the brain, overcoming the blood brain barrier possibly through the olfactory nerve (Oberdörster et al., 2004). These observations raise the question: What are the health effects of inhaled particles on the cardiovascular system and the organs where the particles accumulate?

Epidemiological studies present robust links between exposure to inhaled dusts and cardiopulmonary mortality (Dockery et al., 1993; Samet et al., 2000) but these links are the strongest for individuals with advanced atherosclerotic disease such as those with a previous heart attack (von Klot et al., 2005) or stroke (Zanobetti and Schwartz, 2001). Following an elevation of ambient air pollution, there are increased reports of strokes (Hong et al., 2002; Schwartz, 1994), heart failure exacerbations (Pope et al., 2004), cardiac arrhythmias (Peters et al., 2000), heart attacks (Peters et al., 2004; Pope et al., 2004), and sudden deaths (Schwartz, 1994).

Research findings support two mechanisms that are hypothesized to be involved in the pathological response of the cardiovascular system to air pollution exposure. These mechanisms are inflammation and dysfunction of the autonomic nervous system. The inflammatory mechanism is the concept that particulate material irritates the lungs which results in first a local and then a systemic inflammatory response (Seaton et al., 1995; Seaton et al., 1999). Low grade, chronic inflammation is believed to exacerbate cardiovascular disease (Tranfield et al., 2010). The dysfunction of the autonomic nervous system manifests as irregularities in heart rate, heart rhythm and blood pressure (Peters et al., 2000; Pope et al., 1999).

Extrapolating the cardiovascular effects of inhaled terrestrial dusts to the possible cardiovascular effects of lunar dust on astronauts is fraught with uncertainty because of several critical gaps in the scientific understanding of this topic. To begin, astronauts currently go through an extensive screening process for the purpose of selecting only very healthy individuals. Epidemiologically, the primary correlation between elevated air pollution levels and adverse cardiovascular events is in individuals with known cardiovascular disease (Dockery et al., 1993; Samet et al., 2000). To date little research has focused on the health effects experienced by healthy individuals exposed to air pollution. Virtually all research has focused on very young populations, as well as aging, already ill populations, and astronauts do not fit in either of these categories. Secondly, lunar dust is very different compared to terrestrial dust in terms of chemical composition, method of formation, and exposure to weathering effects and solar irradiation (Horanyi and Stern, 2011; Liu and Taylor, 2011). Furthermore, the site of fine dust deposition in the lungs is more peripheral (Darquenne and Prisk, 2008) and timely removal from the lungs is hypothesised to be reduced in microgravity and most probably in lunar gravity. These variables may have a greater effect on lung toxicity caused by lunar dust than cardiovascular toxicity; however, this is an area that needs to be investigated further before any scientifically sound conclusion can be drawn about how lunar dust will impact astronauts during their time on the lunar surface, and when they return to Earth.

### *5.2. Ocular and skin effects*

Concern about the health effects of lunar dust is dominated by pulmonary considerations, but many lunar operational scenarios have the potential to result in eye exposure and skin exposure to lunar dust (and coarser lunar regolith) (Khan-Mayberry, 2007). For eye exposure, at least two scenarios are envisioned: deposition of airborne particles onto the surface of the eye, and transfer of particles from contaminated objects, such as fingers, that may touch the eye. Exposure of the skin to lunar dust could certainly occur as a result of airborne particles, but operations that involve handling an EVA (extravehicular activity) suit that has been contaminated on the outside with lunar dust, such as EVA suit cleaning and suit repair procedures, likely represent an even greater risk of exposure. Skin exposure to lunar dust may also be of concern if the interior of the EVA suit becomes contaminated with lunar dust (e.g., by donning a suit using contaminated hands), in which case dermal abrasion may take place at sites where the suit rubs against the skin (James and Khan-Mayberry, 2009). The abrasive properties of lunar dust have been documented in a variety of settings (Gaier et al., 2009; Kobrick et al., 2010). Indeed, dermal abrasion due to lunar dust may act in combination with skin damage induced by pressure on the skin at sites of anatomical prominence (finger tips, knuckles, elbows, knees), and may result in breakdown of the stratum corneum, the outermost layer of the skin which functions as the protective barrier of the skin (James and Khan-Mayberry, 2009; Jones et al., 2007).

To address the concern about potential lunar dust effects in the eye, NASA's LADTAG (Lunar Airborne Dust Toxicity Assessment Group) will be sponsoring a study that will be performed with authentic lunar dust from Apollo 14, and will be carried out in two stages. The first stage will consist of an *in vitro* study, using a commercially available testing system that uses differentiated keratinocytes that mimic the architecture of the cornea. To complement the *in vitro* study, Apollo 14 lunar specimens will be tested for eye irritancy using rabbits, to assess possible corneal effects, conjunctival effects, and effects on the iris (Khan-Mayberry, 2007). Examination of lunar dust dermal effects will focus on the risk of skin abrasion. Experiments using authentic Apollo soil specimens and a pig skin abrasion model are currently underway. The ocular studies and dermal studies are expected to reveal information about the biological effects of lunar dust that is complementary to studies of lunar dust pulmonary effects and is operationally relevant for human exploration of the Moon (James and Khan-Mayberry, 2009; Khan-Mayberry, 2007).

## 6. Dust inhalation in animal models

Toxicity testing with particulate air pollutants has been performed extensively both for acute toxicity outcomes as well as chronic toxicity. For diseases such as lung cancer or airway remodelling driven by chronic inflammation, most studies have been done with mice or rats either in whole-body chambers (Heinrich et al., 1995) or in nose-only exposure towers (Pauluhn, 1994, 2008; Wong, 2007). For larger studies with many animal subjects, typical systems are either nose-only tower exposures with 20 animals or more, or multiple whole-body cages exposed continuously to aerosols for up to several hours per day during periods ranging from days up to a year or two (Nikula et al., 1995). While the nose-only towers require a substantially larger work effort compared to the simpler whole body exposure chambers, the dosing precision is considerably better because exposure concentrations are more even, and because the animals are prevented from swallowing pelt-deposited material during grooming. Whole-body cages or nose-only exposures in towers are more suitable for chronic studies requiring a larger number of animals to reach statistical power but are only feasible when the test substance is available in sufficient amounts for production of the exposure atmosphere.

Two common aerosol generator types to feed aerosol into such exposure systems are the Wright dust feeder and the jet mill generator (McClellan and Henderson, 1995), which both consume substantial amounts of test materials. However, the multiple-animal strategy is not optimal if only a limited amount of test substance is available, which is often the case with synthesized or collected air pollution materials (Oberdörster et al., 2005a; Seagrave et al., 2006). Moreover, during single animal exposures it is possible to control exposures of individual animals depending on their ventilation pattern and local aerosol concentration.

So far instillation and insufflation methods have been most common for single animal exposures with limited amounts of test material. In the liquid instillation method, controlled volumes of dissolved or suspended materials are instilled into endotracheally intubated animals (Saffiotti et al., 1968; Warheit et al., 2007). The disadvantage of this method is an uneven distribution of poorly de-agglomerated materials in the lungs, quite different from ambient air exposures in experimental studies or in nature (Driscoll et al., 2000).

The next step towards testing the toxicity of respirable aerosols is the direct insufflation method, where the study material is injected with an air puff directly into the endotracheally intubated lung, either as a liquid spray (Tronde et al., 2002) or as a suspended dry powder formulation (Bosquillon et al., 2004; Larsen and Regal, 2002). With some loss of dosing control, the distribution within the lung is improved, but it is still quite uneven with greater deposition in larger airways compared to more peripheral lung compartments (Bosquillon et al., 2004; Sakagami, 2006). For single-animal rodent exposures to respirable aerosols fewer methods are available. With the modern piezoelectric nebulizers, rats can be exposed to respirable aerosols of dissolved or liquid-dispersed materials (MacLoughlin et al., 2009).

While liquid suspensions can be used for drug delivery without too many problems, this may not be suitable for particulate air pollutants. Most air pollution particles of concern have chemically reactive sites on their surfaces that drive their toxicity. Many of these reactive sites are quenched by water or even humidity, which makes the liquid suspension both an unrealistic and unsuitable mode of administration (Seagrave et al., 2006). Instead, dry powder inhalation is preferred.

For dry powder aerosols, the Dustgun technology offers one possibility to perform respirable aerosol exposures of single lungs that range in size from rodent lungs to dog lungs using small amounts of substance (Gerde et al., 2004; Gerde et al., 2001). The Dustgun is a batch-wise operating generator using compressed air to suspend and deagglomerate dry powders into small volumes of respirable aerosols. The combination of the isolated, ventilated and perfused rat lung (IPL) and the Dustgun aerosol generator has been described elsewhere (Ewing et al., 2006; Selg et al., 2010). Recently, a novel active dosing system was combined with the Dustgun generator, where the measured aerosol concentration and the individual ventilation pattern of the exposed subjects were used to improve the dosing precision (Figure 8). This system can also be used for administering mineral dusts to either endotracheally intubated rodents or nose-only rodent setups. The advantage of exposing intubated instead of nose-only rodents is that substantial losses of exposure material depositing in the nasal airways can be avoided (Sakagami et al., 2003). The disadvantage is that anaesthesia is needed in experiments with tracheally intubated animals.

## 7. Exhaled markers of airway inflammation during spaceflight

### *7.1. Exhaled biomarkers*

Assessing lung health in extreme environments, such as space flight, demands non-invasive techniques. The analysis of biomarkers in the exhaled breath would meet such a requirement. Apart from the atmospheric and classical metabolic gases, the exhaled breath contains a number of other gases in trace amounts, including volatile organic compounds. Some of these compounds can be used to characterize the function of the whole organism (systemic biomarkers), and some are more specifically related to the function of the respiratory system (lung biomarkers). The analysis of breath condensates is an additional non-invasive technique in which breath condensate is collected by cooling or freezing exhaled gas (Kharitonov, 2004; Kharitonov and Barnes, 2001; Patil and Long, 2010; Popov, 2011).

Examples on techniques and associated conditions are:

- exhaled nitric oxide (NO) – airway inflammation (asthma, chronic obstructive pulmonary disease)
- exhaled breath condensate – numerous markers of airway inflammation
- exhaled hydrocarbons – oxidative stress

Of these techniques only the analysis of exhaled nitric oxide is a standardized technique used in clinical practice. Breath condensate and hydrocarbons are both promising techniques but need further development. Additionally, new detection methods and improved analytical techniques of completely novel and potentially more precise markers of airway inflammation might be developed in the future.

*7.2. Exhaled nitric oxide (NO)*

The potential of NO as a marker of airway disease became apparent when Gustafsson et al. (1991) found endogenous NO in the exhalate from both animals and humans. In addition, Alving et al. (1993) and Persson et al. (1994) found increased levels of exhaled NO in asthmatics with ongoing airway inflammation. Normal exhaled NO levels are in the range of 10-35 parts per billion (ppb) in adults, and around 5-25 ppb in children (Taylor et al., 2006). For asthma patients not adequately treated with anti-inflammatory medication, exhaled NO levels are up to 70-100 ppb and sometimes even higher. Exhaled NO is now used as a diagnostic tool in the monitoring of asthma patients on Earth.

Naturally occurring enzymatic processes produce NO from the amino acid L-arginine and oxygen by the enzyme nitric oxide synthase (NOS). There are three different known isoforms of NOS; two constitutive and one inducible (iNOS). The constitutive NOS, i.e., neuronal NOS (nNOS) and endothelial NOS (eNOS), are strictly $Ca^{2+}$-dependent, whereas iNOS is dependent on gene expression regulation.

Potential sources of exhaled NO include the nasal epithelium (Lundberg et al., 1995), the airway epithelium (Asano et al., 1994), the alveolar epithelium (Asano et al., 1994), the vascular endothelium (Ignarro et al., 1987), and the blood (Pawloski et al., 2001). NO enters the airway lumen by gas diffusion driven by a concentration gradient. All three isoforms of the NOS enzymes are present in the lung (Ricciardolo et al., 2004), but normal levels of exhaled NO match only the production quantity from iNOS activation (Ialenti et al., 1993; Lane et al., 2004).

The NO turnover in the lungs is rapid, and part of the NO given off from tissues to the gas phase is exhaled. A similar or even larger quantity is taken up by the blood because NO binds very strongly to the haemoglobin in the blood. Thus, to fully understand the NO turnover in the lungs, the rate of "back diffusion" of NO from its site of formation in the airway wall to the lung alveoli must be assessed, as must the rate of NO transfer from the alveolar space to the blood.

### *7.3. Effects of gravity and atmospheric pressure on exhaled NO*

The atmospheric pressure in future habitats on the Moon and Mars, and in vehicles made for interplanetary space missions, is likely to be substantially lower than that on Earth due to the desire to minimise decompression times prior to extravehicular activity. Studies have shown that normal healthy control values of exhaled NO are reduced in microgravity (Karlsson et al., 2009). Moreover, Hemmingson et al. ((2012)) have shown that pulmonary NO turnover is modified in reduced ambient pressure even though the exhaled NO partial pressure is not changed compared to normal ambient pressure. To establish normative values for exhaled NO in future spaceflight environments, the combined effects of gravity and cabin pressure on exhaled NO will need to be assessed.

## 8. Pulmonary effects

NASA's effort to understand the pulmonary toxicity of lunar dust has been spearheaded by LADTAG. LADTAG has undertaken ground-based animal experiments based on the use of archived lunar soil (Apollo 14) that has been stored in NASA's curation facility (http://www.lpi.usra.edu/captem/). The primary motivation for these experiments has been to derive a permissible exposure limit (PEL) for airborne lunar dust that is in the respirable size range, focusing on particles that are <3 μm in diameter and for which exposure below the PEL is unlikely to cause health effects. The PEL will be used to set requirements for air quality for spacecraft and habitats that can be used in the design of future exploration missions to the Moon.

For these studies (James, 2011; Rask et al., 2012) three different preparations of Apollo 14 lunar dust were used: lunar dust processed in a stainless steel jet mill to reduce particle size to < 3 μm; lunar dust processed by grinding in an oscillating ball mill, using a zirconium oxide grinding jar with zirconium oxide balls; and lunar dust that was not ground by any means, but was processed in a cyclone separator, to select for < 3 μm particles (James, 2011; Rask et al., 2012). Grinding of the Apollo 14 lunar dust specimens was necessary to generate adequate amounts of respirable size lunar dust needed for future rodent inhalation toxicity experiments planned by LADTAG (James, 2011). The three preparations of respirable lunar dust had comparable particle size, as determined by laser diffraction in an aqueous suspension (Rask et al., 2012).

### *8.1. Intratracheal instillation studies*

In a series of experiments performed at the National Institute of Occupational Safety and Health (NIOSH, Morgantown, WV), LADTAG scientists carried out an intratracheal instillation study in rats, using the three Apollo 14 preparations described above, at doses of 1 mg, 2.5 mg and 7.5 mg per rat together with quartz (Min-U-Sil 5) and titanium dioxide as strong responder and low responder controls, respectively (Santana et al., 2010). Using bronchioalveolar lavage fluid cell counts, lactate dehydrogenase levels, albumin, and total protein as markers of inflammation, these experiments demonstrated that the inflammatory response in rats to lunar dust is intermediate between that of the titanium dioxide and quartz controls, at 7 days and at 30 days post exposure. For these bronchioalveolar lavage fluid endpoints, the three different preparations of lunar dust were not distinguishable. Based on the findings of these lavage fluid studies, a preliminary *permissible exposure limit* of 1.0 $mg/m^3$ (6 months, intermittent exposure) was determined, using a benchmark dose analysis method (James, 2011).

*8.2. Differences in toxicity of dust due to method of lunar dust preparation – macrophage reactivity studies*

Freshly-fractured quartz is more toxic than aged quartz, and the leading explanation for this observation is the concept that mechanical disruption of quartz results in the homolytic and heterolytic rupture of the silicon-oxygen bonds ("mechanochemical activation") (Fubini and Hubbard, 2003). Increased surface radical content of the freshly-fractured material is well-correlated with enhanced toxicity. This finding is ascribed to the higher reactivity of dangling bonds and their potential to generate free radicals in aqueous solutions (Fubini, 1998; Fubini and Hubbard, 2003). Is a similar process important for lunar dust, which is exposed to continuous micrometeorite bombardment (Brunetto, 2009) (Figure 9)? The use of three different preparations of Apollo 14 lunar dust in the intratracheal instillation experiments performed by LADTAG provided a means of addressing this question (James, 2011). Indeed, the macrophage reactivity studies suggest that differences in chemical reactivity are in fact correlated with differences in biological response for lunar dust.

Analysis of the three Apollo 14 lunar dust preparations revealed that the material prepared by grinding in a zirconium oxide ball mill exhibited 7-fold greater ability to generate hydroxyl radicals on addition to aqueous buffer, compared to jet-mill processed material or un-ground (cyclone separation only) lunar dust, as judged by the terephthalate assay (Rask et al., 2012). One possible explanation for the observed increased ability to generate hydroxyl radicals in the ball-mill prepared material is an enhanced generation of dangling bonds compared to the other preparation methods. When a degranulation assay (luminol chemiluminescence) was used to analyze macrophages harvested via bronchioalveolar lavage in the LADTAG studies described above, increased macrophage reactivity was observed after exposure to the chemically reactive ball mill-prepared lunar dust, but not after exposure to the other two preparations where less or no mechanochemical activation is expected (Rask et al., 2012). The macrophage reactivity assay used for these studies was the same as the assay used in studies of quartz that showed increased biological responses for freshly ground material versus aged (Porter et al., 2002).

These findings support the concept that lunar dust toxicity is dependent on lunar dust chemical reactivity, and processes that cause mechanochemical activation contribute to this reactivity. Other processes that may cause increased lunar dust chemical reactivity, such as particle radiation exposure or UV light exposure, may be associated with enhanced toxicity as well (Figure 9). Additional ground-based experiments with lunar simulants, lunar dust, and its crystalline and amorphous phases might shed more light on the above findings. *In situ* measurement of the chemical reactivity of lunar dust, as it actually exists on the surface of the Moon, is likely the only way of assessing the net effects of multiple processes that may affect the chemical reactivity of lunar dust (Loftus et al., 2010).

## 9. Knowledge gaps and recommendations

Tables 1 and 2 list the key knowledge gaps that exist for the physical and chemical properties of lunar dust (Table 1) and for its physiological and health effects (Table 2). It must be kept in mind that the samples of lunar material that were brought back from the Moon during the Apollo era were not optimized for studies of their dust component. Moreover, their chemical reactivity is unlikely to mirror that of lunar material *in situ*, since they have not been exposed to radiation and micrometeoroid impacts (Figure 9) for the last 40-50 years – nor have they been kept under vacuum.

Previous work on the toxic effects of lunar material and simulants of lunar material in animal and cellular models has been made with the particles in aqueous media to avoid the large losses of precious material typical of conventional inhalation techniques. New techniques are now being developed which permit more economical inhalation studies in small animals. These techniques include individual and well-controlled administration by nasal or intratracheal routes, the latter being the most exact and economical. Previous work on animal and cellular models has not addressed the effects of dust that has been "activated" with UV or proton radiation. Such studies should be given a high priority in the future. None of the curated lunar samples exist in a pristine state that preserves the surface reactive chemical aspects thought to be present on the lunar surface. Studies using this material therefore carry with them considerable uncertainty in terms of fidelity. As a consequence, *in situ* data on lunar dust properties are required for the ground-based studies to quantify the toxicity of dust exposure during future manned lunar missions.

Based upon the knowledge gaps presented in Tables 1 and 2, we recommend a set of *in situ* measurements on the Moon, and a set of measurements in ground-based, analogue environments. The central goal of the *in situ* measurements is to gain sufficient understanding of the environment on the Moon, and of the physico-chemical features and surface reactivity of lunar dust, so that the toxic effects of lunar dust can be mimicked and studied in ground-based experiments using both re-activated material of lunar origin and lunar dust simulants.

## 10. Summary

Exposure of humans to lunar dust is likely to give rise to toxic reactions and should therefore be avoided, or at least minimized by undertaking a risk assessment and instigating appropriate measures to reduce exposure. Limits for permissible exposure must be based on scientific evidence, which is currently highly incomplete or absent. Dust toxicity is determined by the abundance, size distribution and chemical reactivity of dust particles. Chemical reactivity depends on the composition of dust in interaction with mechanical impacts from meteoroids and the exposure to proton and UV radiation in an environment free of oxygen and humidity. Currently used simulants of lunar dust and existing samples of actual lunar material are useful only for initial pilot experiments with cellular, animal and human experimental models of astronaut exposure to lunar dust. *In situ* determinations of factors such as abundance, size distribution and chemical reactivity of lunar dust particles are essential for a proper assessment of the risks for humans during future manned exploration of the Moon.

## Acknowledgements

The authors comprise the Topical Team on the Toxicity of Lunar Dust (T3LD), and wish to thank the European Space Agency (ESA) for financial support. We thank two anonymous reviewers for suggestions that improved the manuscript. WvW acknowledges funding from the Netherlands Space Office's Principal Investigator Preparatory Programme. DJL acknowledges funding from NASA's Human Research Program, as well as *Center Investment Funds* from NASA Ames Research Center. GKP acknowledges funding from the National Space Biomedical Research Institute (NSBRI) through NASA NCC 9-58.

**Figure captions**

**Figure 1.** (a) Apollo 17 astronaut Gene Cernan covered in dust after extravehicular activity on the lunar surface (photograph courtesy NASA). (b) Size distribution of $n$ = 840 grains recovered from the spacesuit of Apollo 17 astronaut Harrison Schmitt, with a mean of 10.7 µm. Modified from Christoffersen et al. (2009).

**Figure 2.** Steps of cell and tissue interaction with nano and micron-sized particles in the lung. When attained the alveolar space the particle may react with endogenous molecules (step 1). The particle may then be cleared out of the lung either through the mucociliary escalator (step 2) or through alveolar macrophage (AM) clearance (step 3). If reactive, AM activation will follow with release of several factors and recruitment of other immune cells (AM and polymorphonucleate cells, PMN), eventual cell death and establishment of permanent cycles of ingestion (step 4). This process produces chronic inflammation (step 5). Combined with the direct action of the particle (step 6) this will cause damage to the target cells (epithelial, endothelial). If the particle is nano-sized, it may easily escape from the lung to the pleura and to systemic circulation (step 7).

**Figure 3.** The role of particle and cell derived free radicals and reactive oxygen species (ROS) in cell damage, oxidative stress and diseases

**Figure 4.** (a) Cumulative mass distribution of the grain size of selected Apollo soils, modified from Liu and Taylor (2011). (b) Comparison between number distributions of the grain size of the dust fraction of two Apollo samples and of the very fine fraction of lunar dust simulant JSC-1Avf, modified from Park et al. (2008).

**Figure 5.** The effect of bond rupture following mechanical impact or cleavage in crystalline silica, adapted from Fubini et al. (2001). This mechanism can be extrapolated to silicates including lunar dust minerals and amorphous phases, provided most of the chemical bonds are covalent.

**Figure 6.** Total deposition of particles ranging in size from 0.5 to 3 μm in microgravity (panel A), 1 G (panel B) and 1.6 G (panel C). In each case the bars to the left of each pair (with error bars) are experimental data collected in humans. The bars to the right are the predictions based on standard models of aerosol deposition and include the effects of impaction (unshaded portion of the bar), sedimentation (diagonal shading), and diffusion (black filled portion). Reproduced from Darquenne et al. (1997) with permission.

**Figure 7.** Deposition of inhaled aerosol in 1 G (upper line) and in lunar gravity (1/6 G) plotted as a function of volumetric depth in the airways to which the aerosol was inhaled. A penetration volume of 300 ml indicates deposition in the small to medium sized airways, while a penetration of volume of 1200 ml is within the alveoli. Although deposition is reduced in lunar gravity compared to that in 1 G, a given deposition fraction (say 25% as indicated by the dashed line) occurs much more peripherally in reduced gravity. Redrawn from Darquenne et al. (2008) with permission.

**Figure 8.** A schematic drawing of the Dustgun system for the exposures of the anaesthetised and tracheally intubated rat. PN, the pneumotachograph; DG, the DustGun aerosol generator; MF, the end filter; V, the vacuum pump; $\dot{V}_{total}$, the exposure airflow; $\dot{V}_{vent}$, the ventilation airflow, generated by the spontaneously breathing animal; $\dot{V}_{filter}$, the constant component of the exposure airflow. The balance of the airflow streams at the three-way junction is expressed as: $\dot{V}_{total} + \dot{V}_{vent} + \dot{V}_{filter} = 0$.

**Figure 9.** UV light, particle radiation and mechanical disruption caused by micrometeoroid bombardment, individually or in combination, may alter the surface chemistry of lunar dust. Exposure to oxygen and water as well as other atmospheric gases is expected to cause passivation of the chemical reactivity of lunar dust.

**Figure 1**

(a)

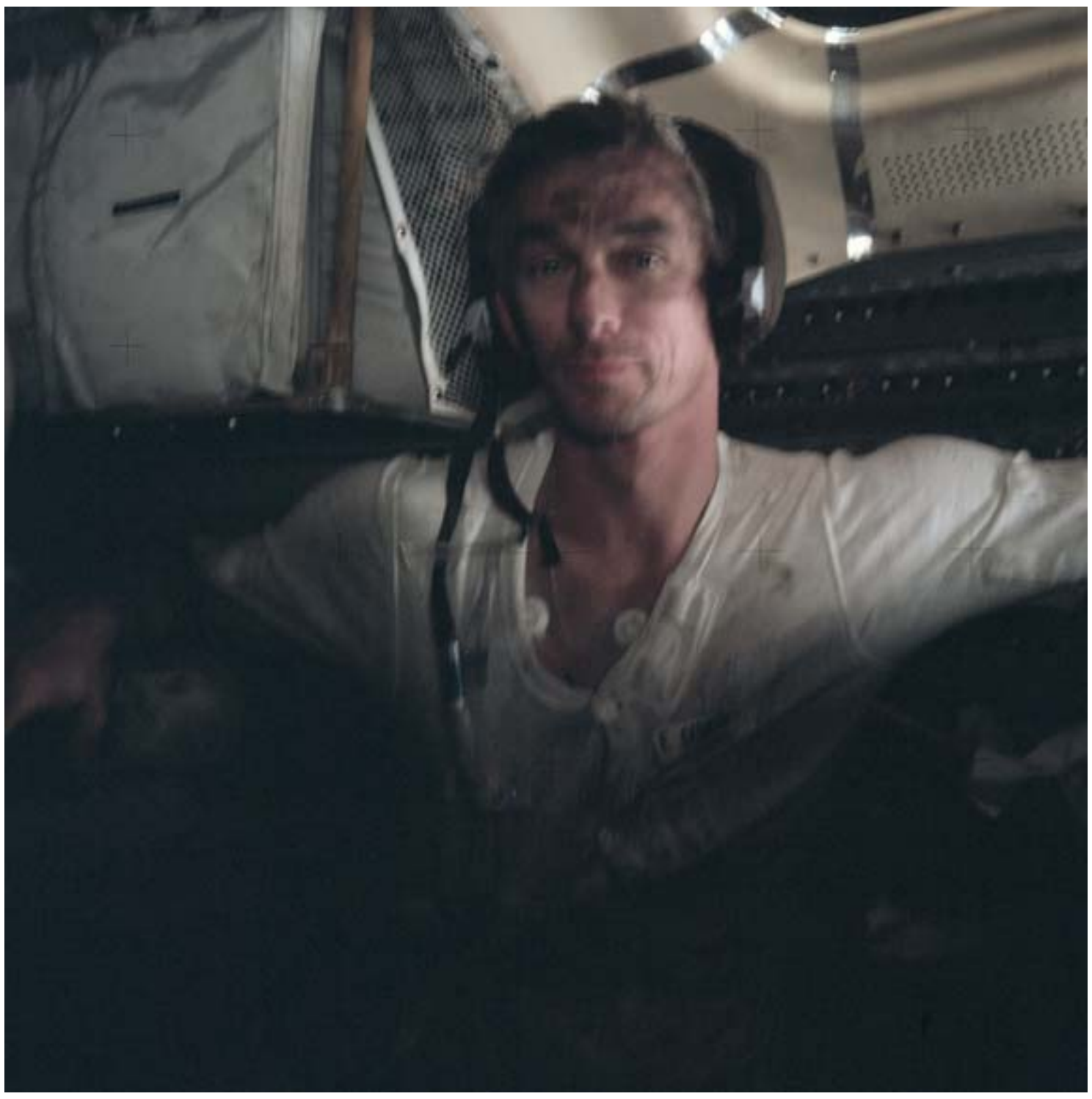

(b)

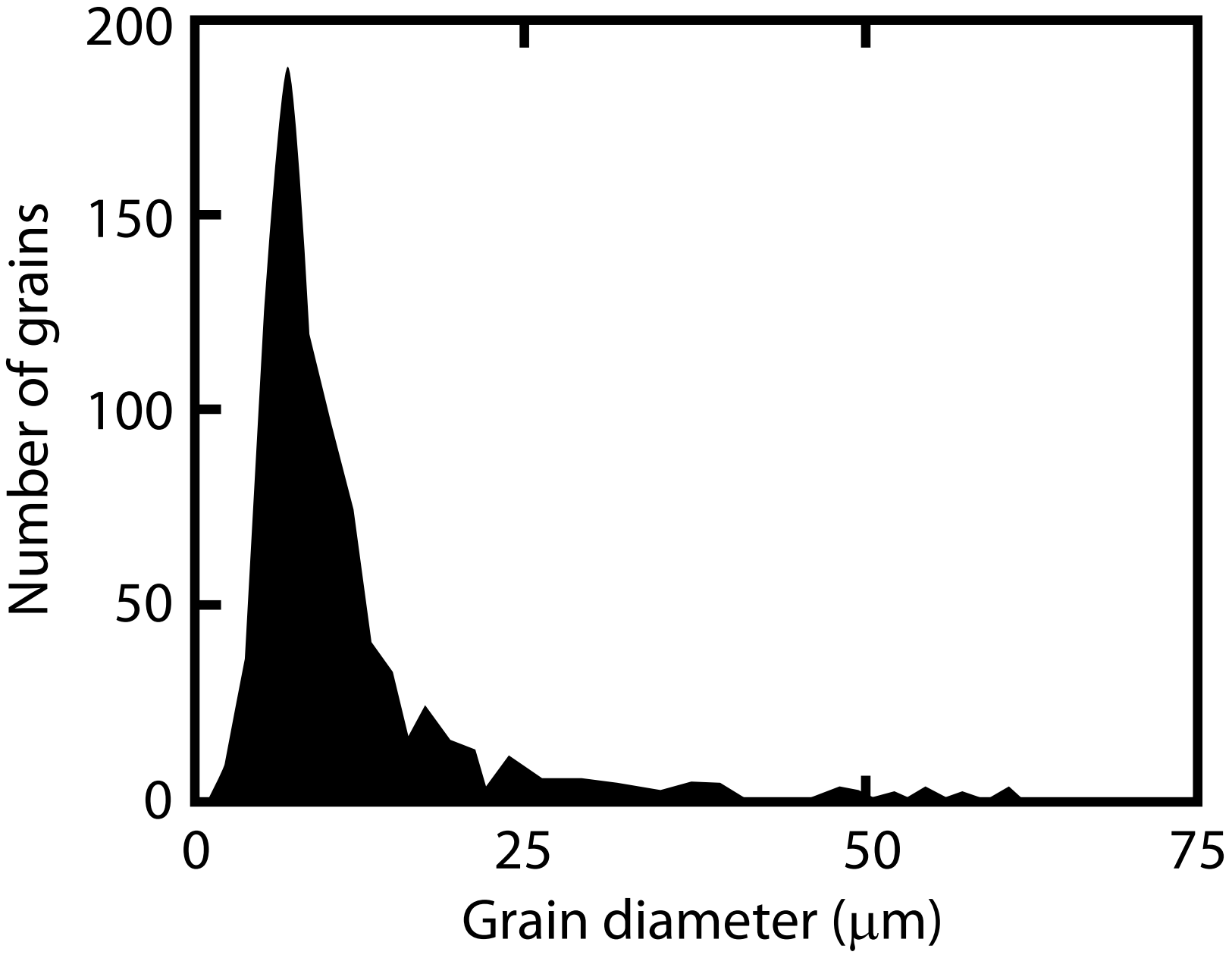

**Figure 2**

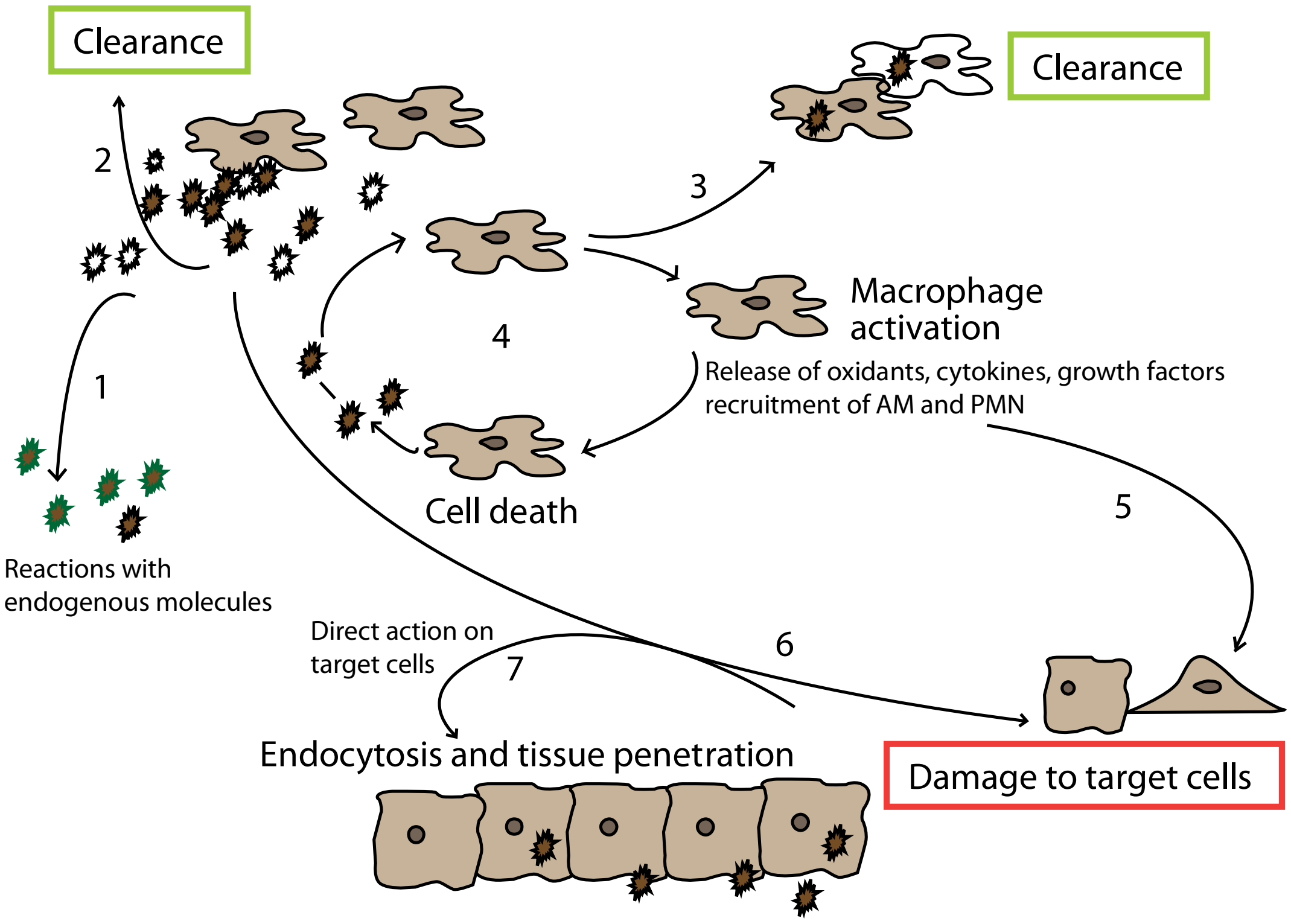

Clearance
2
1
Reactions with endogenous molecules
Clearance
3
4
Macrophage activation
Release of oxidants, cytokines, growth factors recruitment of AM and PMN
Cell death
5
6
Direct action on target cells
7
Endocytosis and tissue penetration
Damage to target cells

**Figure 3**

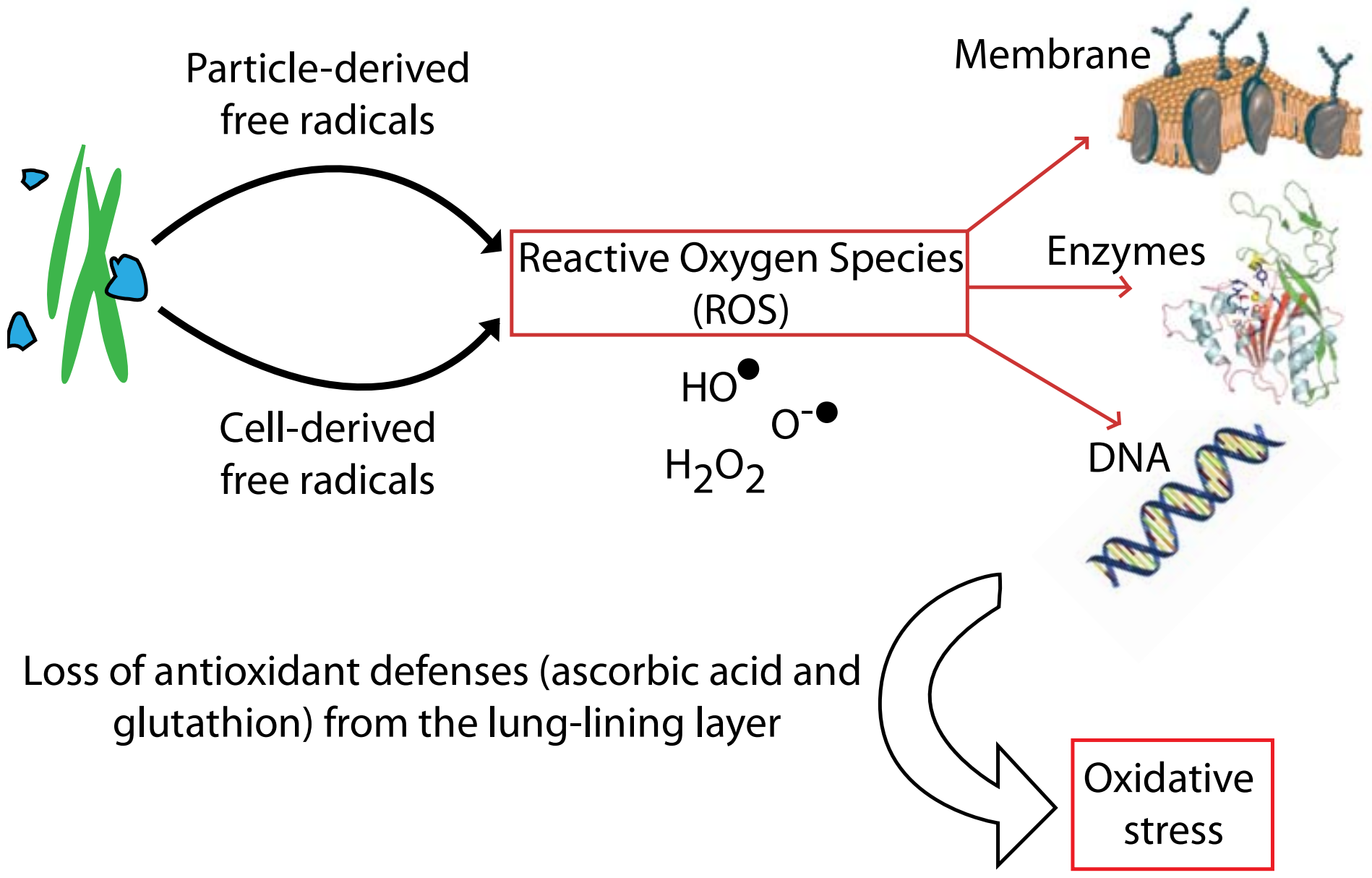

Particle-derived free radicals
Cell-derived free radicals
Reactive Oxygen Species (ROS)
HO•
O-•
$H_2O_2$
Membrane
Enzymes
DNA
Loss of antioxidant defenses (ascorbic acid and glutathion) from the lung-lining layer
Oxidative stress

**Figure 4**

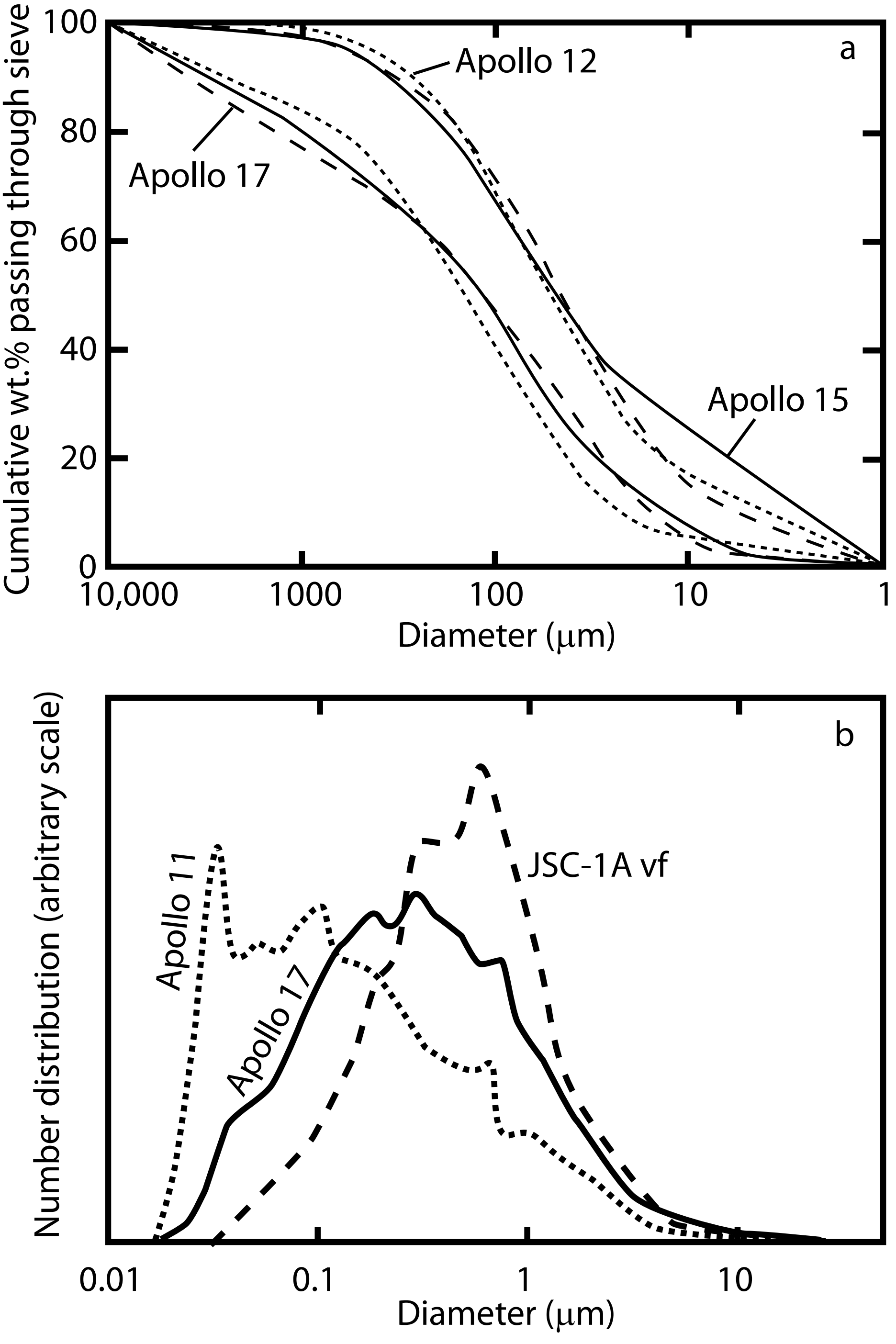

**Figure 5**

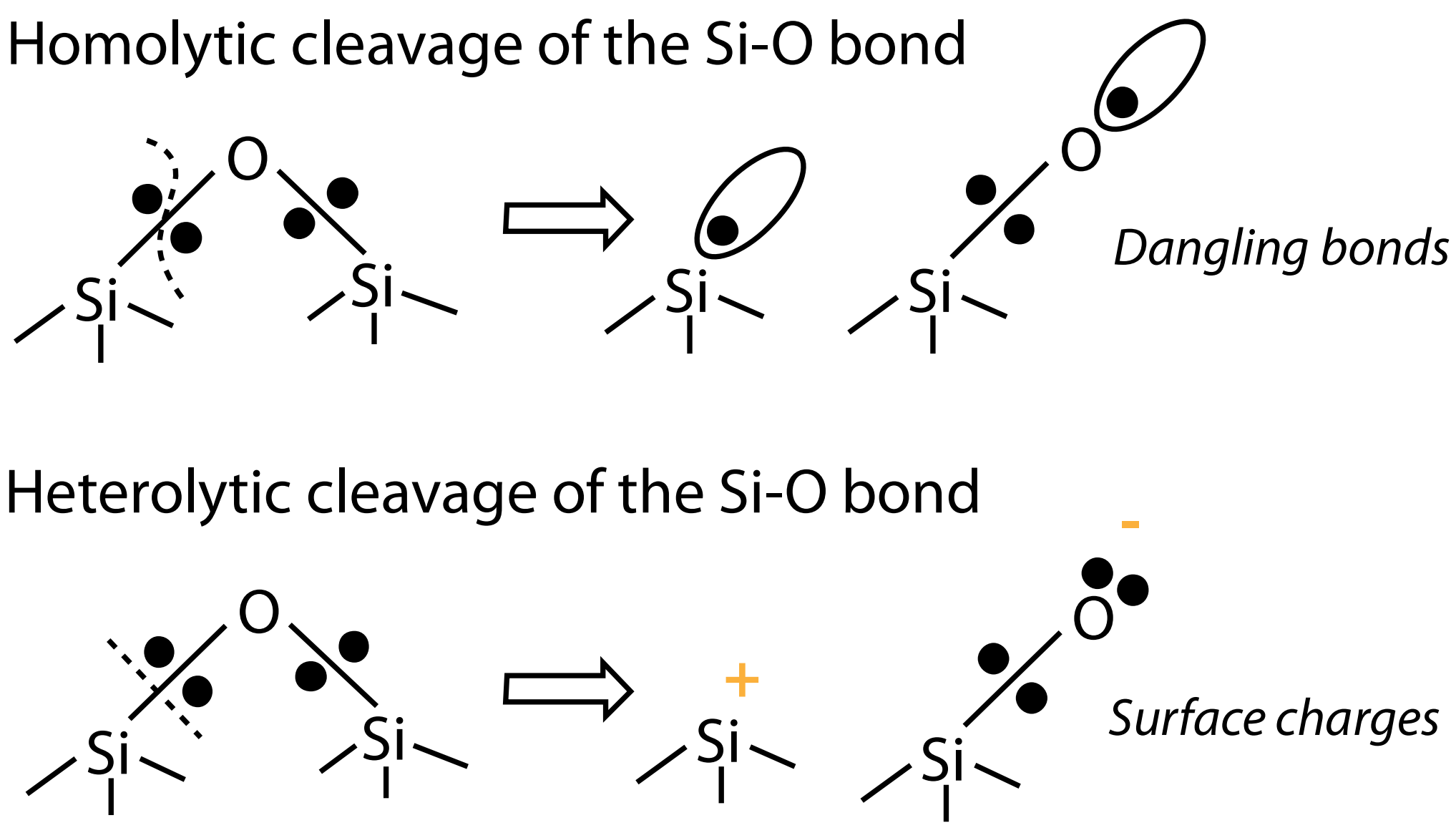

**Figure 6**

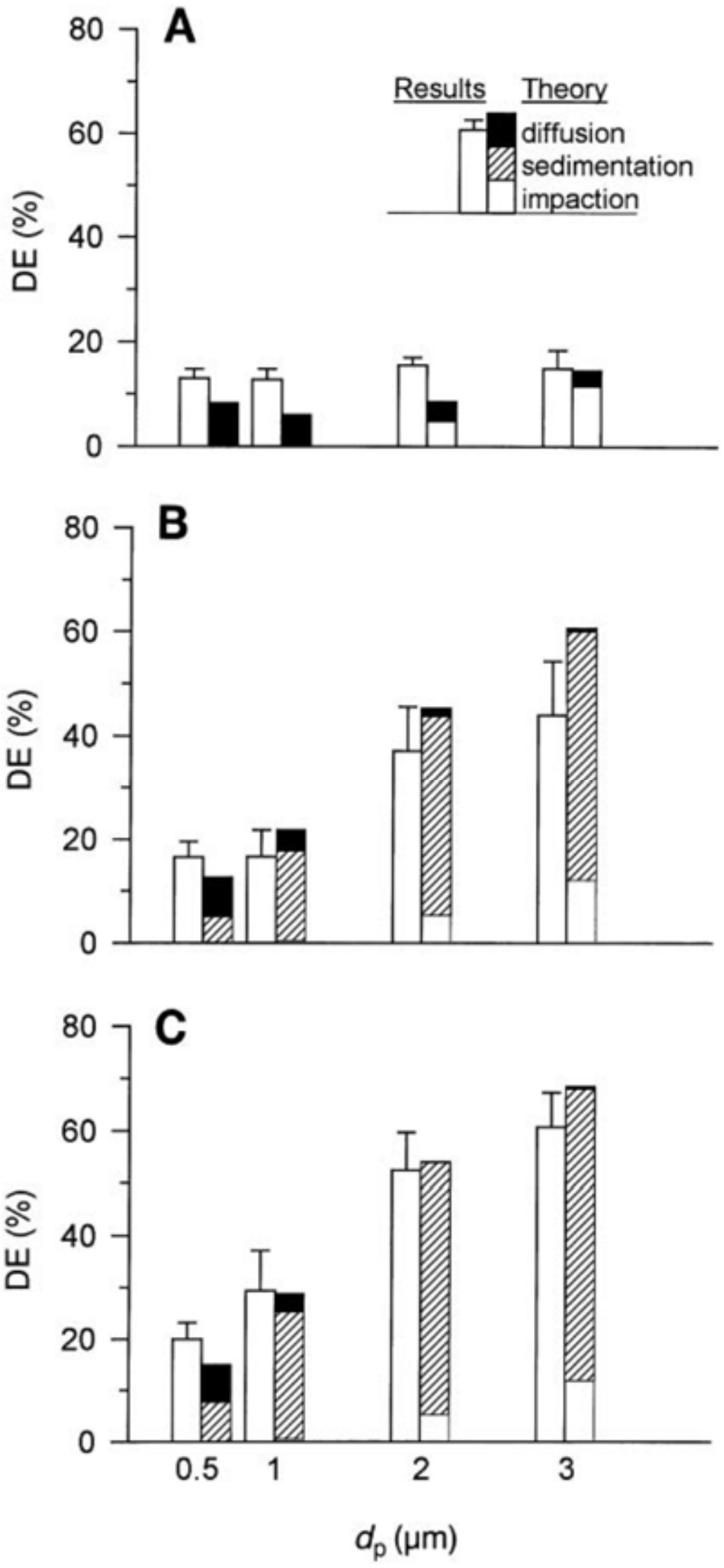

**Figure 7**

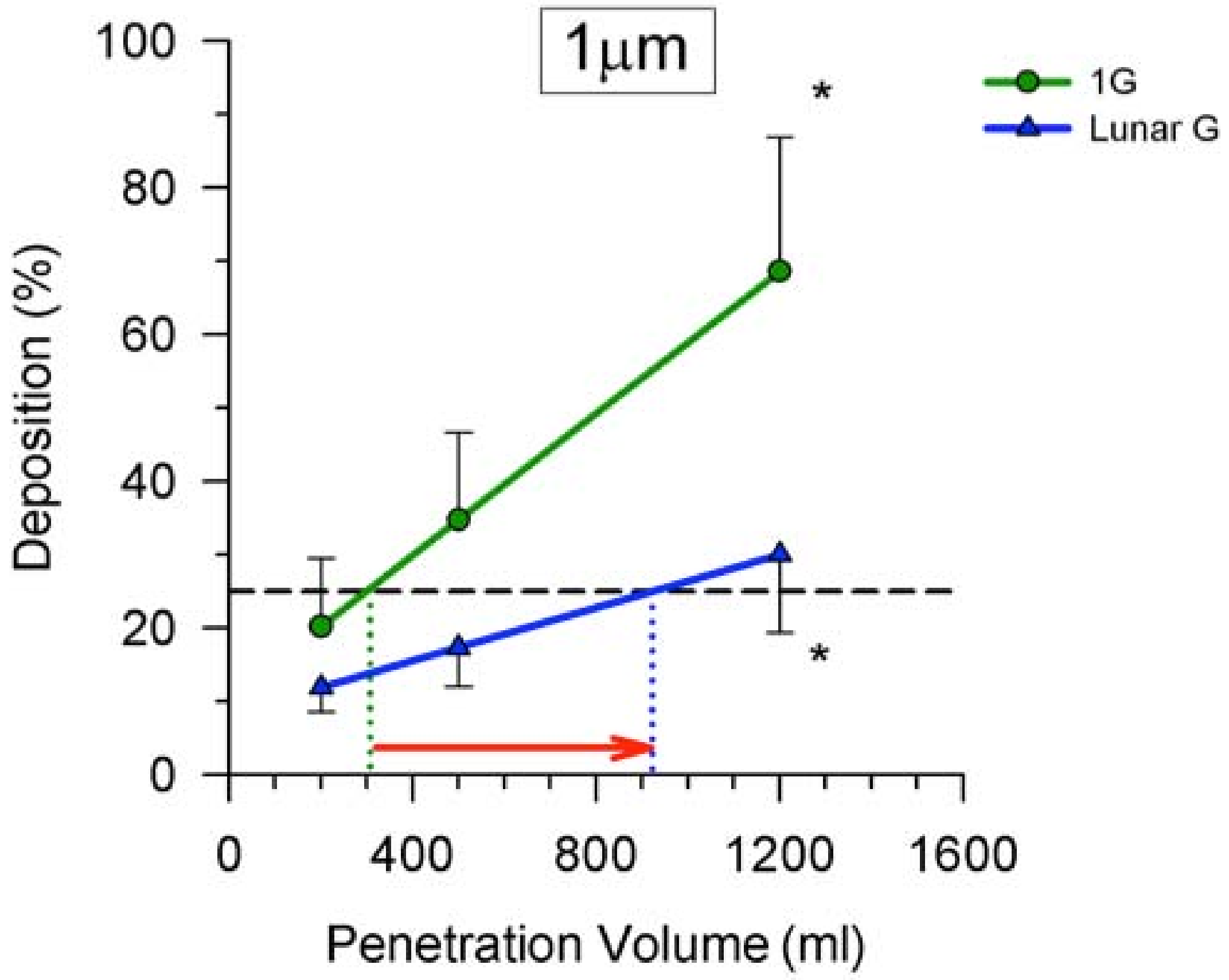

**Figure 8**

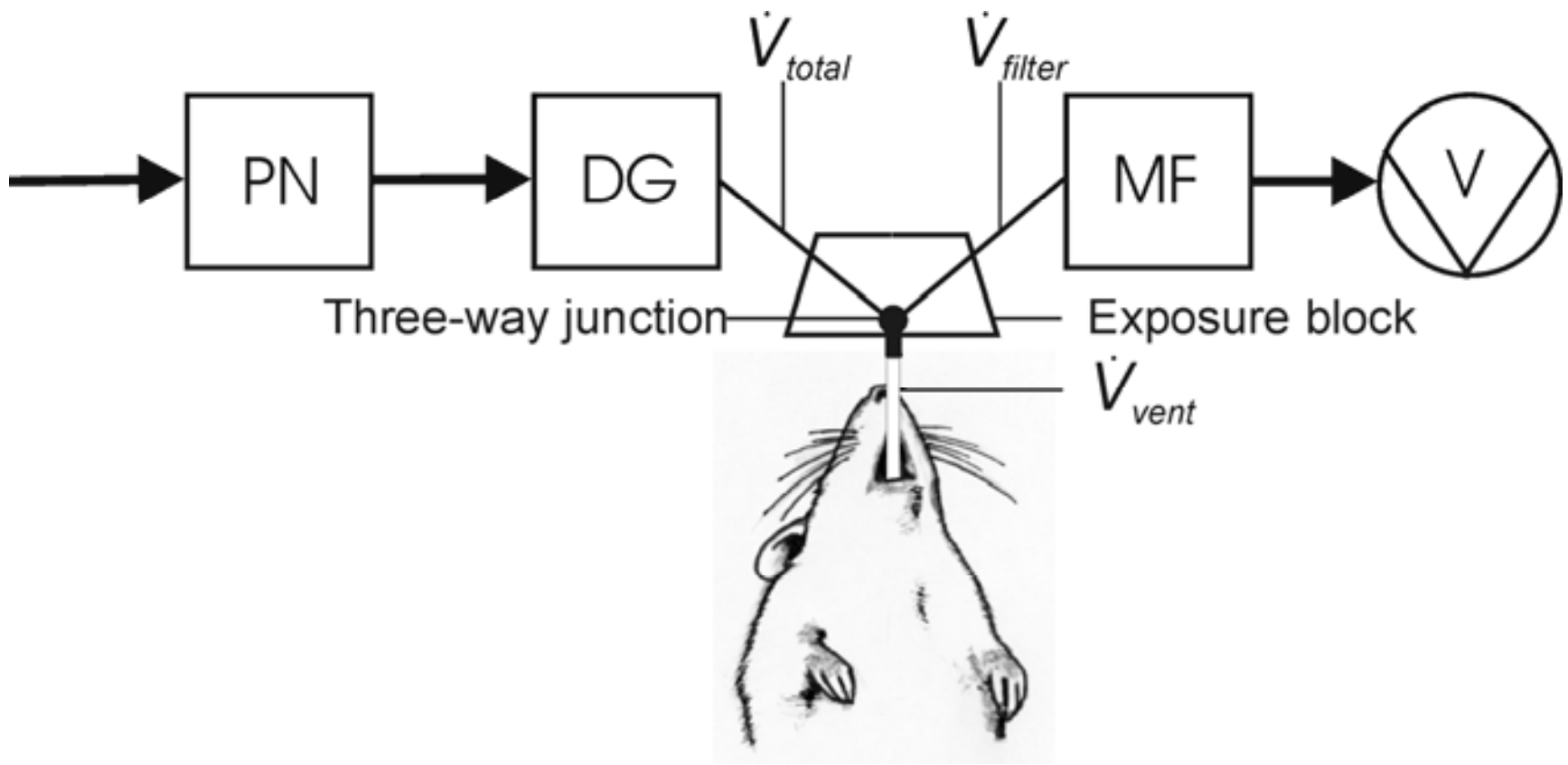

**Figure 9**

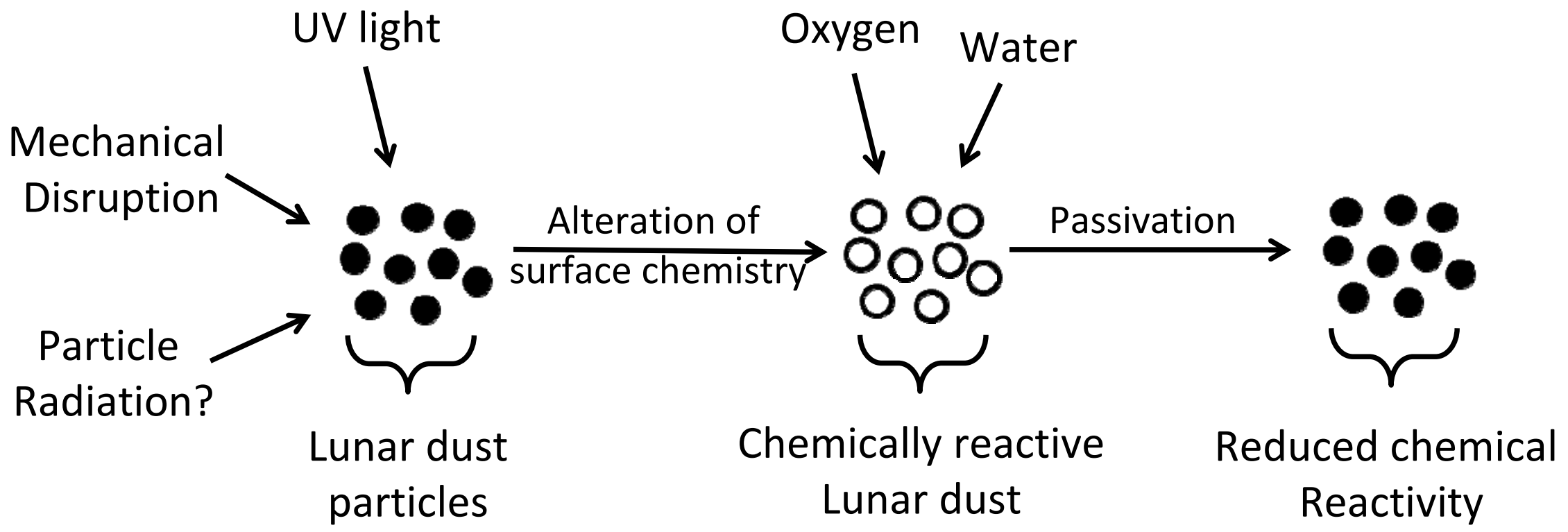

Table 1. Knowledge gaps related to physical and chemical properties of dust in relation to potential toxicity

| *Number* | *Topic* | *Suspected relevance to dust toxicity* | *Current state of knowledge for lunar dust* | *Priority for lunar dust related measurements* | *Type of studies required* |
|---|---|---|---|---|---|
| 1 | Activation of lunar dust due to proton radiation, galactic cosmic radiation, solar particle events, ultraviolet radiation, solar wind implantation, and micrometeoroid impact fragmentation | High | Unknown | High | Ground-based; *In situ* |
| 2 | Identification and evaluation of surface-derived free radical generation pathways | Very high | Limited to some ground-based studies of lunar samples | Critical | Ground-based; *In situ* |
| 3 | Passivation mechanism and rate after exposure of reactive lunar dust to habitat-like atmospheres | High | Unknown | High | Ground-based; *In situ* |
| 4 | Size distribution, specific surface area of the submicron size fraction | High | Uncertain from measurements of lunar samples | High | Ground-based; *In situ* |
| 5 | Size fraction variations of nanophase iron abundance | Very high if it becomes exposed | Abundance known for selected sites and size fractions | High | Ground-based; *In situ* |
| 6 | Efficiency of release of nanophase iron from glassy matrix | High if very efficient | Limited | High | Ground-based |
| 7 | Overall mineralogy | High | Well-known for limited number of sites, rough estimates available from remote sensing | Medium | *In situ* |

| *Number* | *Topic* | *Suspected relevance to dust toxicity* | *Current state of knowledge for lunar dust* | *Priority for lunar dust related measurements* | *Type of studies required* |
|---|---|---|---|---|---|
| 8 | Other physical properties of the smallest size fraction (morphology, dielectric constant, crystallinity) | Medium | Poorly known | Medium | Ground-based;<br>*In situ* |
| 9 | Abundance of volatile components on the lunar surface | Possibly high | Significant uncertainties, particularly for high latitudes | Unknown | Ground-based on pristine Apollo samples;<br>*In situ* |
| 10 | Aerosol generation from lunar dust during typical mission activities | High | Medium from Apollo missions | Medium | *In situ* |

Table 2: Knowledge gaps related to physiology and adverse health effects

| *Number* | *Topic* | *Suspected relevance to dust toxicity* | *Current state of knowledge for lunar dust* | *Priority for lunar dust related measurements* | *Type of studies required* |
|---|---|---|---|---|---|
| 11 | Relation between surface active sites including those generated by radiation and micrometeoroid impact effects and adverse health effects | Very high | Unknown | Critical | Ground-based |
| 12 | Toxicity of nanophase iron | Very high | Unknown | High | Ground-based |
| 13 | Toxicity of ilmenite | Low | Unknown | Low | Ground-based |
| 14 | Dust deposition in sustained low gravity | Medium | Deposition in transient microgravity well-known (aircraft parabolic flights) | Medium | Low-gravity (ISS) |
| 15 | Residence time of particles that deposit in the lung in lunar gravity | High | Currently under investigation | High | Parabolic flights |
| 16 | Rate of mucociliary clearance of particles from the lung in lunar gravity | High | Unknown | High | Suborbital flight |
| 17 | Effects of inhaled lunar dust on the cardiovascular system in disease models | Low | Unknown, based on air pollution research | Low | Ground-based |
| 18 | Effects of inhaled lunar dust on the cardiovascular system in different gravitation environments. | Low | Unknown | Low | Ground-based |
| 19 | Effects of inhaled lunar dust simulant in an animal model | High | Extrapolation from instillation of particles with an aqueous vehicle | High | Ground-based |
| 20 | Effects of inhaled “activated” simulant (Proton, UV radiation) in an animal model | High | Unknown | High | Ground-based |

| *Number* | *Topic* | *Suspected relevance to dust toxicity* | *Current state of knowledge for lunar dust* | *Priority for lunar dust related measurements* | *Type of studies required* |
|---|---|---|---|---|---|
| 21 | Effects of inhaled “activated” Apollo era lunar material | High | Unknown | High | Ground-based |
| 22 | Time course of deactivation after irradiation of simulants and actual lunar material | High | Unknown | High | Ground-based |
| 23 | Effects of combined microgravity and reduced pressure on the NO turnover in the lungs of humans | High | Partly known from micro-gravity and altitude separately | High | Low-gravity (ISS) |
| 24 | Effects of dust-induced airway inflammation on exhaled markers other than NO | High | Unknown | High | Ground-based, animal and human experiments |